\documentstyle[psfig,prl,aps,epsf]{revtex}
\begin{document}
\twocolumn[\hsize\textwidth\columnwidth\hsize\csname
@twocolumnfalse\endcsname
\title{
Beyond Hebb: Exclusive-OR and Biological Learning$^*$ 
}   
\author{Konstantin Klemm$^\dagger$, Stefan Bornholdt, and Heinz Georg Schuster}
\address{Institut f\"ur Theoretische Physik, Universit\"at Kiel, 
Leibnizstrasse 15, D-24098 Kiel, Germany
\\
$^\dagger$Current address: The Niels Bohr Institute, Blegdamsvej 17, 
DK-2100 Copenhagen, Denmark 
\\
$^*$ published as Phys.\ Rev.\ Lett. 84 (2000) 3013. } 
\maketitle
\date{today} 
\begin{abstract}
A learning algorithm for multilayer neural networks 
based on biologically plausible mechanisms is studied.
Motivated by findings in experimental neurobiology, 
we consider synaptic averaging in the induction of plasticity changes, 
which happen on a slower time scale than firing dynamics. 
This mechanism is shown to enable learning of the exclusive-OR (XOR) 
problem without the aid of error backpropagation, as well as to 
increase robustness of learning in the presence of noise. 
\medskip \\
\end{abstract}
]

Since the early days of neurophysiology we have 
evidence of biological mechanisms serving as the basis 
for learning and information processing in the brain.    
Cajal's pictures showing networks of intertwined 
nerve cells readily lead to the hypothesis of information 
flow and processing in these networks \cite{Cajal11}. 
Subsequently formulated theoretical models of the neuron,
as by McCulloch and Pitts \cite{MCPitts43}, and  
the Hebbian learning rule, postulating synaptic 
strengthening for simultaneous pre- and postsynaptic 
activity \cite{Hebb49}, sparked the development of 
algorithms for neuronal learning and memory. 
The development of learning algorithms, however, 
took place almost decoupled from biological validation, 
partly due to lack of detailed knowledge of the 
neurophysiology of learning, but also due to their 
success in applied fields (``connectionism'', 
``machine learning''). 
Among the first models were layered assemblies 
of formal neurons (Perceptrons) combined with gradient 
rules defining the synaptic weights \cite{Rosenblatt62}. 
Later, combining Hebb's strictly local rule with 
symmetrically connected formal neurons defined the 
Hopfield model of simple associative learning 
\cite{Hopfield88}. However, only a complicated 
non-local learning rule, now known as error backpropagation,  
finally was able to solve simple non-linear learning problems 
as the learning of the exclusive-or (XOR) function \cite{Backprop86}. 
This complicated form of reverse information transfer, 
however, has not been observed in biological circuits
\cite{ZipserAndersen88}. 

For computation in biological nervous systems 
the question remains, which underlying biological 
processes are capable of the most general form of 
learning \cite{Crick89}, including problems of the XOR class.  
A more biologically plausible learning concept is 
learning by reinforcement and recently a number of models 
along this line have been formulated \cite{Reinforce96,SuttonBarto98}. 
One such model by Barto and Anandan combines a local 
mechanism of synaptic plasticity changes with a global 
feedback signal indicating information worth memorizing 
\cite{BartoAnandan85,Barto85}. A remaining problem in 
these models is the regulation of mean activity level 
in large networks which has been attacked 
by Alstr\o m and Stassinopoulos \cite{AlstrStass95} 
and Stassinopoulos and Bak \cite{StassBak95,Bak96}.  
An even more elegant mechanism has been proposed by 
Chialvo and Bak \cite{ChialvoBak99} 
with reinforcement through negative feedback which is  
motivated by the observed long-term depression (LTD) in
biological networks. In this algorithm, the dynamics 
of synaptic plasticity comes to a halt when learning 
reaches its goal, just by definition. 
While we think that this is a very interesting 
approach to formulating a biologically plausible 
learning mechanism, this model suffers a severe 
restriction in learning capabilities. 
It has been shown to work well on simple tasks as 
non-overlapping pattern sets, however, it is not 
able to learn tasks as the XOR function, 
at least not without unreasonably large numbers 
of neurons and very long learning times.  
It is, therefore, nearly as limited as the early 
single layer perceptron models that, for this reason, 
nearly paralyzed the research in neural networks 
in the seventies (mainly following the sobering 
analysis of perceptron capabilities by Minsky 
and Papert \cite{MinskyPapert69,AndersenRosenfeld88}). 

In the following we will study a model in this spirit 
which, however, does not exhibit this restriction. 
Let us first define the model, then report numerical 
results on its learning capabilities. We will then 
discuss the robustness of our model in the presence 
of noise. The letter concludes with a discussion of 
the motivation of our model from current findings 
in neurobiology.  

Consider a layered network of binary formal neurons 
$x_i\in\{0,1\}$, with $I$ input sites $x_0, 
\dots, x_{I-1}$, $J$ hidden sites $x_I, \dots, 
x_{I+J-1}$, and $K$ output units $x_{I+J}, 
\dots, x_{I+J+K-1}$. The adjacent 
layers are completely connected by weights $w_{ji}$ 
from each input to each hidden unit and from each hidden 
unit to each output unit. 
In addition, each weight is assigned an internal 
degree of freedom, acknowledging the finite time 
scale of synaptic plasticity induction 
as will be discussed below.   
In the model this is represented by an additional   
discrete variable $c_{ji}$ associated to each weight
$w_{ji}$ serving as a synaptic memory during learning.

The network dynamics is defined by the following steps.  
The input sites are activated with external stimuli 
$x_0,\dots,x_{I-1}$.
Each hidden node $j$ then receives a weighted input 
$h_j=\sum_{i=0}^{I-1} w_{ji} x_i$. 
Its state is chosen according to a probabilistic rule 
s.t.\ each hidden neuron fires with 
probability $p_j = a^{-1} \exp(\beta h_j)$ with
the normalization $a = \sum_j \exp(\beta h_j)$.
We consider the low activity limit of the network
where only one hidden neuron fires at a time.  
Each output neuron $k$ now receives an input sum 
$h_k=\sum_{j=I}^{I+J-1} w_{kj} x_j$ with the only 
non-zero contribution from the firing hidden neuron 
$j^\ast$ such that $h_k=w_{kj^\ast}$. The above
probabilistic rule applies to the output layer as well, 
determining one firing output neuron $x_{k^\ast}$  
which represents the output of the network corresponding 
to a given input pattern. 
Note that in the low activity limit used here,  
the probabilistic rule is a stochastic approximation of the 
winner-take-all rule \cite{WTA}. We think our 
variant based on local dynamics is biologically 
more realistic than supplying global information 
of which neuron has the highest input sum within a layer.  
In the limit $\beta \rightarrow \infty$, 
the neuronal activity in our model follows exact 
winner-take-all dynamics, since then $\max_j p_j= a^{-1}
\exp(\beta \max_j h_j) \rightarrow 1$. This deterministic case
has been used in the network model of Chialvo and Bak
\cite{ChialvoBak99}. Here, however, we consider stochastic
models with finite values of $\beta$.

Now it remains to specify the learning dynamics of the 
network weights $w_{ji}$ themselves. For each activation 
pattern, the network output is compared to the target 
output and a feedback signal $r$ returned to the network, 
with $r=+1$, 
if its output neurons represent the predefined target output,
given the current input,
and $r=-1$ otherwise. Depending on this binary feedback, 
connections and corresponding counter values are updated. 
All ``active'' synapses $w$ (and corresponding counter 
values $c$) for which pre- and postsynaptic sites have 
been simultaneously active are updated as follows.  
The feedback signal is subtracted from the memory 
$c$ of each active synapse according to: 
\begin{equation} \label{eq:updatec}
c \rightarrow c^\prime = \left\{   
\begin{array}{llrcl}
\Theta, & {\rm if}\, &             & c-r & >\Theta \quad (\ast) \\
c-r,    & {\rm if}\, & \Theta \geq & c-r & \geq 0\\
0,      & {\rm if}\, & 0         > & c-r. &       \\
\end{array} \right.
\end{equation}
Thus, each counter $c$ is an error account of the 
corresponding synapse. The capacity of the account is 
given by the memory size $\Theta$. In case this 
threshold is exceeded [marked by $(\ast)$ in 
equation (\ref{eq:updatec})] the synapse is penalized, 
i.~e., it is weakened by a constant amount $\delta$:  
\begin{equation}
w \rightarrow w^\prime = w - \delta.  
\end{equation}
(Alternatively, a multiplicative penalty 
combined with a constant growth of weights 
has been successfully checked, too.)  
Therefore, the counter averages over the record of
a synapse, instead of penalizing each single error 
at the moment it occurs. Note that the model by 
Chialvo and Bak \cite{ChialvoBak99} is just this 
latter case and is obtained by setting 
$\Theta=0$ and $\beta=\infty$. 
After these changes to weights and counters
the learning cycle is iterated by presenting 
another---possibly different---pattern of 
stimuli to the network. 

Note that $\beta$ and $\delta$ are not independent
parameters; changing the value of $\delta$ does not
affect the dynamics, as long as the product $\beta \delta$
is kept constant and the weights are rescaled correspondingly.
Furthermore, the firing probabilities are conserved under
transformations that add the same value to all outgoing
connections of one neuron. We could therefore keep the values
of the weights in a bounded domain without changing the model
dynamics.

\begin{figure}[thb]
\let\picnaturalsize=N
\def\picsize{85mm} 
\def\picfilename{./fig1.eps}
\let\epsfloaded=Y
\centerline{\ifx\picnaturalsize N\epsfxsize \picsize\fi
\epsfbox{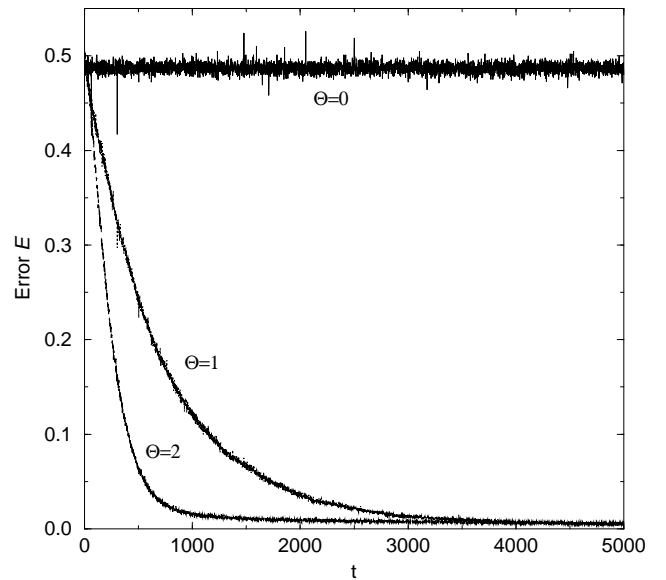}}
\caption{Learning curves show the effect of the internal 
synaptic memory under weak noise, ($\beta=10$). In the 
case of synapses without internal memory ($\Theta=0$)
the error remains close to 0.5, practically no adaptation 
to the desired function (XOR) takes place. However, 
networks with one-step memory synapses ($\Theta=1$) 
quickly reduce the residual error, indicating a fast 
adaption process. Increasing the memory length 
($\Theta=2$) leads to even more efficient learning. Each 
learning curve is an ensemble average of 10000 independent 
runs. 
\label{fig:lcurves}
}
\end{figure}

Let us next demonstrate the learning capability 
and robustness of the model by simulating 
an XOR learning task. 
The network used has $I=3$ input sites $x_0$, 
$x_1$, and $x_2$, with the input site $x_0\equiv 1$
serving as bias. 
The hidden layer has $J=3$ neurons, the minimum 
number necessary to represent the XOR function 
in the present architecture. $K=2$ output 
neurons represent the two possible outcomes with 
only one of them active at a time. 
The initial values of the 
weights $w$ are uniformly chosen random numbers 
$\in [0,1]$, all counters $c$ are set to $0$, and $\delta=1$. 
The four patterns 
of the XOR function are presented with equal probability. 
Fig.\ \ref{fig:lcurves} shows learning curves for 
memory sizes $\Theta=0,1,2$ with $\beta=10$
and averaged over 10000 independent runs each.   
The displayed error $E$ is the fraction of simulation 
runs that have produced an incorrect output at the 
considered time step. We find that learning takes place 
with $\Theta\geq1$ only, where the error quickly 
converges to zero, whereas with $\Theta=0$ 
as in the model of Chialvo and Bak \cite{ChialvoBak99}  
no learning 
takes place at all. The error remains constant hardly 
below the ``default'' of $0.5$ (this holds for the 
whole simulation time of 100,000 trials, not shown here).

The obviously dramatic effect of the synaptic memory can 
be understood in the following way: Any synapse that is 
involved in failure---meaning that pre- and postsynaptic 
firings have occured prior to unsuccessful output of the 
network---is a candidate for decrement. In the case
$\Theta=0$ all such ``failing'' synapses are weakened, 
such that on repeated presentation of the same stimuli the 
activity is likely to be lead to a different output neuron. 
This is a simple and reasonable principle as long as our 
learning goal is the mapping of just one pattern of stimuli 
or a set of non-overlapping patterns. However, the task of 
learning a non-trival logical operation as the one we are 
facing here, requires a more elaborate mechanism:
The immediate weakening of all synapses, that are involved 
in failure for a certain pattern, eventually destroys a 
useful structure for the successful mapping of other patterns. 
This is avoided by the synaptic memory considered here: 
Only if a synapse is repeatedly involved in failure, 
its efficacy is reduced. 

The idea of averaging over errors and updating the weights 
on a slower time scale than sample presentation is well known 
from batch learning methods \cite{Heskes96}. 
In those methods, errors are determined over a whole sweep 
through the pattern set and subsequently weights are updated 
synchronously. However, those algorithms fail to explain 
learning in biological neural systems as they rely on biologically 
implausible mechanisms as, for example, back-propagating errors. 
In fact, what we wish to define here is a learning method based 
on purely local dynamics, where weight changes are based only on 
information that is locally available (the two adjacent neurons 
of a synapse) with nothing more than a single global reinforcement 
signal---exactly the information that is available to a  
biological synapse. A first step in this direction would be 
a trivial ``localized'' version of batch learning where weight 
changes are based on the global reinforcement signal, only. 
Indeed, this works for single layer networks, however, 
fails for learning XOR-type problems in multi-layer networks. 
Here, our work proposes a solution, using a synaptic error 
account combined with asynchronous updating of the synaptic weights. 
It can be viewed as a generalization of the Hebbian learning rule: 
While the Hebb rule alone is not able to make a network learn the 
XOR, the above extension does so. The resulting network is a 
self-contained dynamical system with local dynamical rules 
defined in a way that the overall network dynamics results 
in adaptive learning of general logical functions including 
the XOR problem. Besides learning XOR as shown here, 
the algorithm also proved to learn logical functions 
of higher dimensions and complexity.  

\begin{figure}[thb]
\let\picnaturalsize=N
\def\picsize{85mm} 
\def\picfilename{./fig2.eps}
\let\epsfloaded=Y
\centerline{\ifx\picnaturalsize N\epsfxsize \picsize\fi
\epsfbox{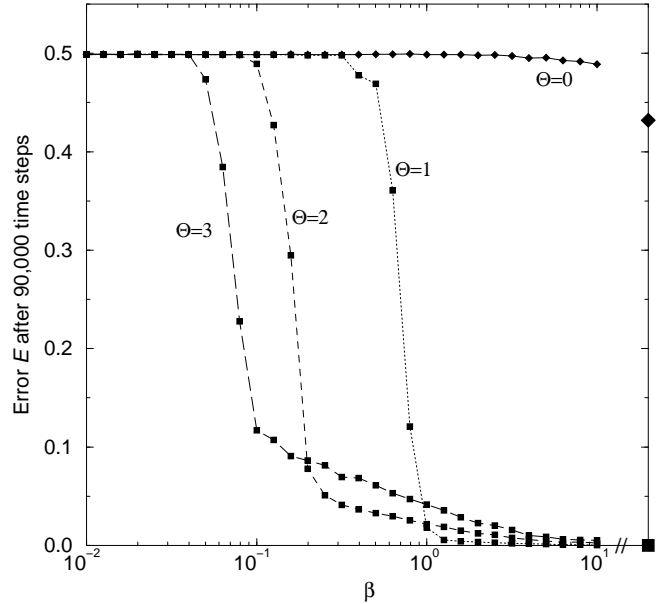}}
\caption{Longer synaptic memory allows for learning in noisier 
networks. The critical value of $\beta$ for the transition
between the non-learning (high error) and the learning 
(low error) regime decreases with the memory length $\Theta$
with larger memory meaning higher robustness against noise.
On the right vertical axis the residual errors for deterministic
networks ($\beta=\infty$) are shown. Without synaptic memory
($\Theta=0$, diamond; this corresponds to the model of Chialvo 
and Bak, see text) we still find a high error, otherwise
($\Theta=1,2,3$, large square) complete learning is achieved.
Displayed errors are averages over time steps 90,000
to 100,000 of 100 independent simulation runs.
\label{fig:transitions}
}
\end{figure}

The aspect of protecting synapses from too quick changes 
has further implications with regard to the network's 
robustness against noise. Fig.\ \ref{fig:transitions} 
demonstrates the effect of the inverse noise 
level $\beta$ on the mean residual error after 90,000 trials. 
For fixed memory length $\Theta$ we find a sharp transition 
from a regime of non-learning, characterized by $E=0.5$, 
to a regime of effective learning with $E\rightarrow0$.
We conclude that the network is able to learn just as long
as the information gain provided by the feedback signal is
larger than the information loss caused by the uncertainty 
of the stochastic neural dynamics. The effect of increasing 
the memory length $\Theta$ is obvious: The critical point 
between the two regimes is shifted to lower values of $\beta$, 
i.~e., higher noise. Synapses with larger memory can average 
out the uncertainty and therefore enable stochastically 
firing networks to adapt to their environment.

Now let us briefly discuss the biological motivations 
for the choice of mechanisms used in the model above. 
First, observations in experimental neurobiology show clear 
evidence that modulation of long-term potentiation 
(LTP) and depression (LTD) via external signals occurs 
(i.e., modulation of plasticity of weights).   
In one example from the hippocampus CA1 region,
which is involved in learning and memory formation, 
modulation mediated by dopamine has been verified 
\cite{Dopamine98}. In particular, when dopamine is applied 
during or shortly after LTD activity, one observes that 
LTD is suppressed (and LTP can appear instead). Learning 
activity can thereby receive feedback via dopamine
which then modulates synaptic plasticity, in particular LTD. 
Indeed, hormone signals are widely known to interfere 
with learning and memory formation. For example adrenal 
hormones have been shown to enhance susceptibility 
for LTD \cite{Coussens97}, an effect which has even 
been found following behavioral stress in living animals 
\cite{Kim}. A broad class of other factors that modulate 
synaptic plasticity have been classified, sometimes 
summarized as ``metaplasticity'' \cite{metaplasticity}. 
We believe that further research in this area will provide 
new insights in the computational mechanics of biological 
nervous systems.  

Furthermore, progress has been made in exploring 
the mechanisms of retrograde feedback in LTP and LTD. 
Evidence accumulates in favor of some physiological mechanisms that 
feed back the postsynaptic activity to the presynaptic site. 
A possible mechanism recently proposed for LTD is 
the messenger nitric oxide evoking a specific presynaptic 
biochemical cascade which, eventually, interacts with 
the intracellular mechanisms for vesicle formation and 
loading \cite{NO}. The subsequently reduced transmitter 
release establishes a long term depression of this 
synaptic pathway. An interesting observation is the 
long time scale of this process of the order 
of 15 minutes \cite{NO}, especially when compared to that of neuronal 
firing packages. This opens up the possibility that considerable 
time averaging may occur in the course of inducing LTD. 
The effect of such a synaptic averaging on learning has been 
simulated above by an internal counter associated with 
each synaptic weight. Further experimental research 
on the timing of externally induced LTD and the lifetimes 
of the biochemical agents involved in the retrograde 
signaling cascade may show to what extent synaptic 
averaging in the induction of plasticity changes is 
established in nature.  

To summarize, we studied a biologically motivated model 
for goal-directed learning in multilayer neural networks. 
In contrast to existing models, synaptic plasticity is 
based on a time-averaged individual failure rate of each 
synapse. Thereby, learning of general logical functions 
(including XOR) is possible on the basis of local synaptic 
plasticity alone, combined with homogeneous failure feedback. 
In particular, no error backpropagation is needed.  
The presented algorithm also works in the presence
of noise, where internal errors are compensated for by 
the averaging of each synapse: only persistent failure is punished.

\end{document}